\documentclass[twocolumn,showkeys,preprintnumbers,amsmath,amssymb]{revtex4} 

\usepackage{graphicx}
\usepackage{dcolumn}
\usepackage{bm}

\begin{document}

\title{Mesoscopic Quantum Thermo-mechanics: a new frontier of experimental physics}
\author{E. Collin$^{*,\dag}$}

\address{(*)  
Institut N\'eel - CNRS UPR2940, 
25 rue des Martyrs, BP 166, 38042 Grenoble Cedex 9, France
}

\date{\today}

\begin{abstract}
Within the last decade, experimentalists have demonstrated their impressive ability to control mechanical modes within mesoscopic objects down to the quantum level: it is now possible to create mechanical {\it Fock} states, 
to entangle mechanical modes from distinct objects, 
store quantum information 
or transfer it from one quantum bit to another, 
among the many possibilities found in today's literature. Indeed mechanics {\it is} quantum, very much like spins or electromagnetic degrees of freedom.
And all of this is in particular referred to as a new engineering resource for quantum technologies. 
But there is also much more beyond this utilitarian aspect: invoking the original discussions of Braginsky and Caves where a quantum oscillator is thought of as a quantum detector for a classical field, namely a gravitational wave, 
it is also a unique sensing capability for {\it quantum fields}.
The subject of study is then {\it the baths} to which the mechanical mode is coupled to, let them be known or {\it unknown} in nature. 
This Letter is about this new potentiality, that addresses stochastic thermodynamics, 
potentially down to its quantum version, 
the search for a fundamental underlying (random) field postulated in recent theories that can be affiliated to the class of the Wave-function Collapse models, 
and more generally open questions of Condensed Matter like the actual nature of the elusive (and ubiquitous) Two-Level Systems present within all mechanical objects. 
But such research turns out to be much more demanding than the usage of a few quantum mechanical modes: all the known baths have to be identified, experiments have to be conducted in-equilibrium, and the word ``mechanics'' needs to be justified by a real ability to move substantially the centre-of-mass when a proper drive tone is applied to the system. 
\end{abstract}


\maketitle

\section{Introduction}

Motion plays a very specific role in Quantum Mechanics (QM) \cite{leggett}.
In the first place, it is at the heart of the definition of {\it heat}: solid-state phonons are nothing but quasi-particles constructed from the quantized motion of real particles. Throughout the paper, phonons will thus be referred to as motion energy quanta traveling in the bulk. When dealing with a {\it localized collective mechanical displacement}, we will use the terminology {\it mechanical mode} (like in a phononic crystal, a levitating sphere, a beam or drum structure). 
The manuscript is focusing on a very specific type of experiments, where a mesoscopic mechanical object is passively cooled to such low temperatures that all the modes are in their quantum ground state, in equilibrium with their environment. With a proper quantum-limited driving and detecting scheme (that should be discussed in due time below), new possibilities are at reach which address three scientific fields that we shall first introduce. \\

{\it Thermal properties - }
Mesoscopic moving objects are perfectly suited to the study of {\it classical stochastic thermodynamics} \cite{stochasticC}.
In small systems, {\it fluctuations} of thermodynamic quantities can be as large (or even larger) than their mean values,
which leads to {\it individual thermodynamic trajectories} that can hurt our intuition; 
but they nonetheless follow well-defined statistical laws that extend the macroscopic version of the second law 
 of thermodynamics \cite{jarzinski}.
More strikingly, in mesoscopic systems the experimentalist has the ability to record information about these individual thermodynamic realizations, and feed it back to the device in order to ``rectify'' them at will: thus selecting specific events that lead to an apparent violation of thermodynamics laws, implementing an experimental realization of a {\it Maxwell's demon} \cite{maxwelldemon}.
Paradoxes are lifted by considering the thermodynamics of the whole apparatus, namely the system under study {\it plus} the feed-back control (the demon). The loss of entropy on the system's side is thus compensated by an increase of the demon's one, who used information and ``destroyed it''. 
Beautiful experiments have been realized using {\it mechanical model systems} like trapped colloidal particles \cite{ciliberto}, DNA strands \cite{DNA} or high-$Q$ cantilevers \cite{LudoPRL}. For instance, the {\it Landauer erasure principle} has been verified in such systems, proving that erasing a bit of information produces a minimal amount of heat of $k_B T \, \ln[2]$ (when the information is lost in a bath at temperature $T$) \cite{ciliberto,LudoPRL}. 

Stochastic thermodynamics has also been addressed in experiments involving electrons and photons; namely {\it electromagnetic degrees of freedom} \cite{cilibertoEleq,JukkaElec}. 
With a Single-Electron Box (SEB) whose charge is recorded with a Single-Electron Transistor (SET), the large fluctuations predicted by stochastic thermodynamics have been measured and fit to theory \cite{saira}. 
Feed-back control has also been implemented, with a computer playing the role of the demon and adjusting a gate electrode in real-time by utilizing the SET acquired data \cite{koski}. In similar setups, thermodynamic cycles have been implemented which demonstrated work extraction up to 66$~\%$ of $k_B T$ (while on average, no work is produced)  \cite{OliveJukka}.
Bringing down such experiments to low enough temperatures, it would be possible in principle to address {\it quantum stochastic thermodynamics} \cite{stochasticQ,JukkaElec,auffeves}. Even though it requires a good deal of technology, electromagnetic degrees of freedom in mesoscopic systems are routinely brought to the quantum regime in a variety of experiments performed around the world. 

Quantum thermodynamics opens up new (almost philosophical) questions linked to our old paradoxes based on demon's monitoring thermodynamic trajectories; 
 a quantum calorimeter {\it does affect} the system under study, and this has to be fundamentally taken into account \cite{stochasticQ,auffeves}.
What shall be the impact of the measurement protocol, and of quantum-coherence and entanglement (which are all key features of QM, see discussion thereafter)?
How to take into account {\it zero-point-fluctuations} (ZPF)? 
Expressing the {\it fluctuation-dissipation theorem} for heat, one finds a deviation to the famous classical formula: energy fluctuations in a mesoscopic system is not only related to the heat conductance towards the baths, there is also a quantum correction that arises from the ZPF of all 
degrees of freedom 
 \cite{averinjukka}. This theoretical prediction appears as a finite-frequency energy noise term that remains at $T =0\,$K, but which can be neglected at high temperature. The specificity here is that temperature {\it is not} itself 
a degree of freedom; the fluctuation-dissipation theorem applied to a single bosonic mechanical mode does not show any such correction, even in the limit $T \rightarrow 0\,$K.

With no surprise for the reader, quantum aspects are far from trivial. In the famous {\it input-output theory} which is the basis for quantum information processing,  
the ZPF of the localized light/microwave modes (of optical cavities or $RLC$ circuits) are actually {\it maintained} by the zero-point-fluctuations of the traveling fields it couples to \cite{inputoutput}. This reasoning applies equally well to mechanics (at least to the knowledge of the author), 
which implies that the ZPF of the mechanical mode are guaranteed by the {\it quantum noise} entering from the baths it couples to.
These baths shall thus already attract our attention; and are actually at the core of the present discussion.
Experimentally, nothing has been done yet on truly quantum aspects (even the SEB-based experiments are still in the classical regime) \cite{stochasticQ}. 
And more specifically, the mechanical model systems used so far are clearly not suited for this purpose. 
In this manuscript, we shall discuss these issues within the framework of a dedicated experimental platform.

Electric currents are quantized in mesoscopic conductors at low temperatures \cite{landauer,buttiker,ElecQuantumexperiment}. 
It has been measured in the first place using 2 Dimensional Electron Gases (2DEGs) with quantum point contacts, demonstrating that the electric current is transported by conduction channels, which support each a {\it quantum of electric conductance} $e^2/h$.
Similarly, heat transport is also 
quantized in a universal way at $T \rightarrow 0\,$K: the conductance reduces to $\pi^2 k_B^2 T/(3h)$ per channel, regardless of  the type (statistics) of particles \cite{rego,kryve,theoryQcond,theoryMiles}.
Experimentally, this has been implemented using 
 photons \cite{photonsheat,timofeev}, and non-interacting electrons in a single ballistic conduction channel \cite{electronheat}. 
In the latter case, the Wiedemann-Franz law has been verified down to the single-channel limit \cite{wiedermanFranz}.
Even with abelian anyons, quasi-particles emerging in the fractional quantum hall regime (and having a statistics in between Bosons and Fermions), the validity of the thermal quantum is assessed \cite{anyons}.

For phonons, {\it one experimental article} reports directly on the measurement of the quantum of thermal conductance \cite{schwabconduct}.
This work was a tour-de-force realized in the year 2000, implying very sensitive measurement techniques, reliable cryogenics (down to milliKelvins), plus a great deal of nanofabrication. 
Another experiment supports this finding \cite{thermalANC}; however other attempts using improved technologies and devices seem to disprove it \cite{oliveconduct,stochasticQ}.
It turns out that the more recent results on thermal conductance of dielectric nano-bridges demonstrate a thermal conductance {\it much lower} than the expected quantum; the quantum of thermal conductance seems to be only an upper limit, reached with an {\it ideal} transmission coefficient (equal to 1) between the baths and the thermal conductor \cite{theoryMiles}.
In practice, this coefficient seems to be usually much less than 1; 
and it even seems to be temperature-dependent (the low-temperature conductance {\it is not} linear-in-$T$) \cite{oliveconduct}. This has been speculatively attributed to Two-Level-Systems (TLS) present in the dielectrics, a domain of Material Science that shall be addressed below. Some theoretical work has been devoted to the actual calculation of the transmission coefficient, including experimental imperfections (namely here surface scattering) \cite{SanCross}; 
but to our knowledge degrees of freedom {\it internal to the constriction zone} have never been taken into account.
Obviously, one has to conclude that not much is known about phonon transport (and related fluctuations) in the quantum regime. \\

{\it Material Science - }
Motion is also a unique tool to probe the constitutive materials of mechanical objects.
A main issue in solid-state low temperature physics is the understanding of the intimate nature of amorphous matter.  
Its properties are understood in the framework of the Two-Level-Systems model \cite{TLS,anderson}: microscopic entities (atoms or groups of atoms) can switch positions (both by thermal activation and quantum tunneling), with a rather flat density of states (with respect to TLS microscopic properties) due to their broad variety.
These TLSs can be constitutive of the dielectric in use (as in amorphous Silicon Oxide, Silicon Nitride), or can be defects of the structure (in the sense that 
the material does naturally present a (small) structurally disordered region that hosts TLSs, as for instance the thin naturally grown Aluminum Oxide layer of Al-based superconducting circuits). 
In almost all systems where their imprint appears, their {\it actual microscopic nature} is unknown. This triggers intensive research for both fundamental reasons (glasses deserve to be understood in their own right) and technological reasons \cite{enssTLS}:
TLSs are ubiquitous in mesoscopic devices, and limit their performances, as can be illustrated by the tremendous work which has been devoted to the improvement of coherence times of superconducting quantum bits (qubits) \cite{martinis}.

The low-temperature mechanical properties of mesoscopic structures are understood in terms of the TLS theory \cite{NEMSdamping}.
All structures including high-frequency phononic crystals \cite{TLSfreqshift} and nanotube beams \cite{nanotubedampingVdZ} display the same characteristic features: a mechanical frequency shift that follows a logarithmic temperature-dependence, and a mechanical damping that shows a $T^\alpha$ power law (plus a constant attributed to clamping losses \cite{clamps}). 
In principle, this power law is related to the dimensionality of the mechanical object (3D,2D,1D), and the actual TLS-bath coupling mechanism (towards phonons or electrons); the slope of the frequency logarithmic shift can also be related to the damping \cite{TLS}. 
This self-consistency check is not very much discussed in the micro/nano-mechanics literature, apart from specific cases \cite{TLSolive}.
The way mechanics probes TLSs is {\it fundamentally different} from electronic devices like qubits \cite{ustinovTLS} and superconducting cavities \cite{natcommTLS}: for the latter, the coupling is due to the dielectric moment that (some) TLSs carry, while for the former the mechanism is a distortion of the TLS potetial due to strain, which affects {\it all} TLSs.

The mechanical properties in the milliKelvin regime of many different mesoscopic devices have been reported, see e.g. Ref. \cite{NEMSdamping} for a review.
But only few works present a thorough fit to TLS theory, discussing explicitly the TLS-coupling mechanisms to the baths \cite{TLSfreqshift,TLSolive}. 
As for electronic systems, intrinsic sources of frequency {\it and damping} fluctuations drastically limit the capabilities of mesoscopic mechanical systems  \cite{oliveNoise,natnanotechsansa}. These sources are usually attributed to the seemingly unavoidable presence of TLSs, but to date no clear picture has emerged.
Theoretically, the interaction of a mechanical mode with a few TLS has been investigated \cite{remus}. 
Exploiting explicitly the quantumness of intrinsic TLSs coupled to a mechanical system is an exciting new possibility that has emerged with ground-state cooled mechanical modes. 
The mechanical resonant coupling to a TLS would extend (to strain) the experimental knowledge of what has been achieved (electrically) with qubits \cite{ustinovTLS}.
 Beyond potential applications like storage of quantum information, which would be the mechanical analog of what has been realized with spins and qubits \cite{bertet}, such experiments would directly give access to the quantum properties of {\it single} TLSs, potentially revealing their nature. \\

{\it Foundations of QM - }
Motion is intimately linked to the foundations of Quantum Mechanics \cite{leggett}.
Despite its successes in describing the atomistic world, QM is (obviously) {\it not} complete \cite{bassiReview}.
On a pure theoretical level, the quantum formalism is not compatible (in its current state) with General Relativity (GR): 
space-time is treated as a given {\it classical} framework, while in GR the {\it metric} is defined from the mass distribution. 
This means that a proper (grand-unified) quantum-gravity theory should take into account that the (fluctuating) quantum fields that describe matter influence the space-time frame itself.  
As such, some attempts have been made to create a superior theory where space and time are raised at the level of quantum operators, see e.g. \cite{bassiReview,singh2011}. 
Even though these GR concepts are intimately linked to what follows (and this shall be reminded whenever appropriate), a theoretical discussion is clearly outside of the scope of the present manuscript.

On a more pragmatic level, there is a long-standing issue about the quantum-to-classical transition: what is known as {\it the measurement problem}.
Let us first recall the accepted so-called Copenhagen interpretation of QM, which perfectly efficiently {\it describes} quantum dynamics (and all we need to know to compute measurement expectations). A thorough discussion of these QM concepts can be found e.g. in Refs. \cite{zurek,weinberg}.
Physical measurable properties are described by observables, namely Hermitian operators: the possible values that can be taken by these properties are the (real valued) eigenvalues of these operators. The eigenstates of these operators generate a Hilbert space which represents 
all accessible information. 
A generic quantum state is a linear combination of state vectors in the Hilbert space describing the problem at hand. This Hilbert space is obtained by the product of the system-under-study subspace $\mathcal{S}$ (i.e. the system's observables), and the environment subspace $\mathcal{E}$ (i.e. all the rest, which includes our measuring apparatus).
This space is usually very large, and contains states that have {\it no analog} in classical mechanics: superposition states (i.e. while classically a bit can only be 0 or 1, a quantum bit can be any linear combination), and entangled states (states constructed from vectors of $\mathcal{S}$ and $\mathcal{E}$ that cannot be factorized out; the system is intimately linked to the environment by construction).
The dynamics followed by the quantum state is perfectly deterministic: it is a unitary (and as such reversible) evolution derived from the Hamiltonian of the problem (the basic principle of quantum dynamics, equivalent to the Schr\"odinger equation). 
Starting from a generic quantum superposition, the interactions of our system-under-study with degrees of freedom from $\mathcal{E}$ (which are described by a term in the Hamiltonian that combines observables from the 2 subspaces) will make this state {\it decohere}: 
after a short period of time, only a given (small) class of states survives, which are linear superpositions of so-called pointer-states. This phenomenon has also been named einselection: the environment (including the measuring apparatus) ``selects'' states that are eligible for a measurement, and destroys any coherence between them \cite{zurek}. In other words, the entanglement with the environment leads to an apparent irreversible evolution for the sub-system under study, by ``diluting'' information in the very large subspace $\mathcal{E}$. These states are then stable under the Hamiltonian evolution, and are the grounds for classicality. 

But the measurement process has not taken place yet. 
The weight of each pointer-state in the quantum superposition obtained after decoherence is understood as a probability of obtaining this specific read-out from the measuring apparatus: this is known as the {\it Born rule}. In the Copenhagen description, this is a specific axiom that is added to the QM framework: the final step of any measurement is a {\it projection} onto one of the pointer states, that shall then evolve again according to the Hamiltonian dynamics \cite{bassiReview,zurek}. 
This projection, also called wave-packet reduction, is fundamentally different from the Hamiltonian evolution (it is both non-unitary, and non-deterministic) and is essentially at the source of all famous paradoxes of quantum mechanics. Even if ``the method works'', this is clearly unsatisfactory on the philosophical level. Quoting 
S. Weinberg: ``we ought to take seriously the possibility of finding some more satisfactory other theory, to which quantum mechanics is only a
good approximation'' \cite{weinberg}.

As already realized by J. von Neumann in the early years of QM, the process of wave-packet collapse cannot be built by a sole unitary evolution \cite{neumann}.
Since then, physicists have been continuously addressing the problem, trying to justify this axiom by a more fundamental phenomenon; and one way to resolve the issue so that collapse does indeed occur is to invoke a {\it nonlinear correction} to the idealized (linear) Schr\"odinger's equation, together with a {\it source of intrinsic noise} of some kind \cite{bassiReview}. As was already pointed out when discussing issues of general relativity, space-time has to be involved to some extent;
and {\it position operators} as building blocks for nonlinear terms ensure that {\it any macroscopic object} shall be affected \cite{bassiReview}. 
Such (yet phenomenological) theories have to fulfill strict requirements: on a  higher level, their structure should preserve basic concepts like energy conservation, and on a lower level they have to be negligible in the microscopic world while guaranteeing classicality at the macroscopic scale. 
There is a plethora of such models; we refer the interested reader to Ref. \cite{bassiReview}.
Some of these models consider explicitly the quantum field of {\it gravitons} as the main source of collapse for mechanical macroscopic states \cite{penrose,stamp,DiosiG}.
But the broadest class of theories do not make assumptions on the fundamental nature of the stochastic field, and simply introduce characteristic lengthscales $\lambda_C$ and (single-nucleon) rates $\Gamma_C$ for the collapse to occur: these are Continuous Spontaneous Localization models (CSL) \cite{bassiReview,SCL}.

		\begin{figure}
		\centering
	\includegraphics[width=14cm]{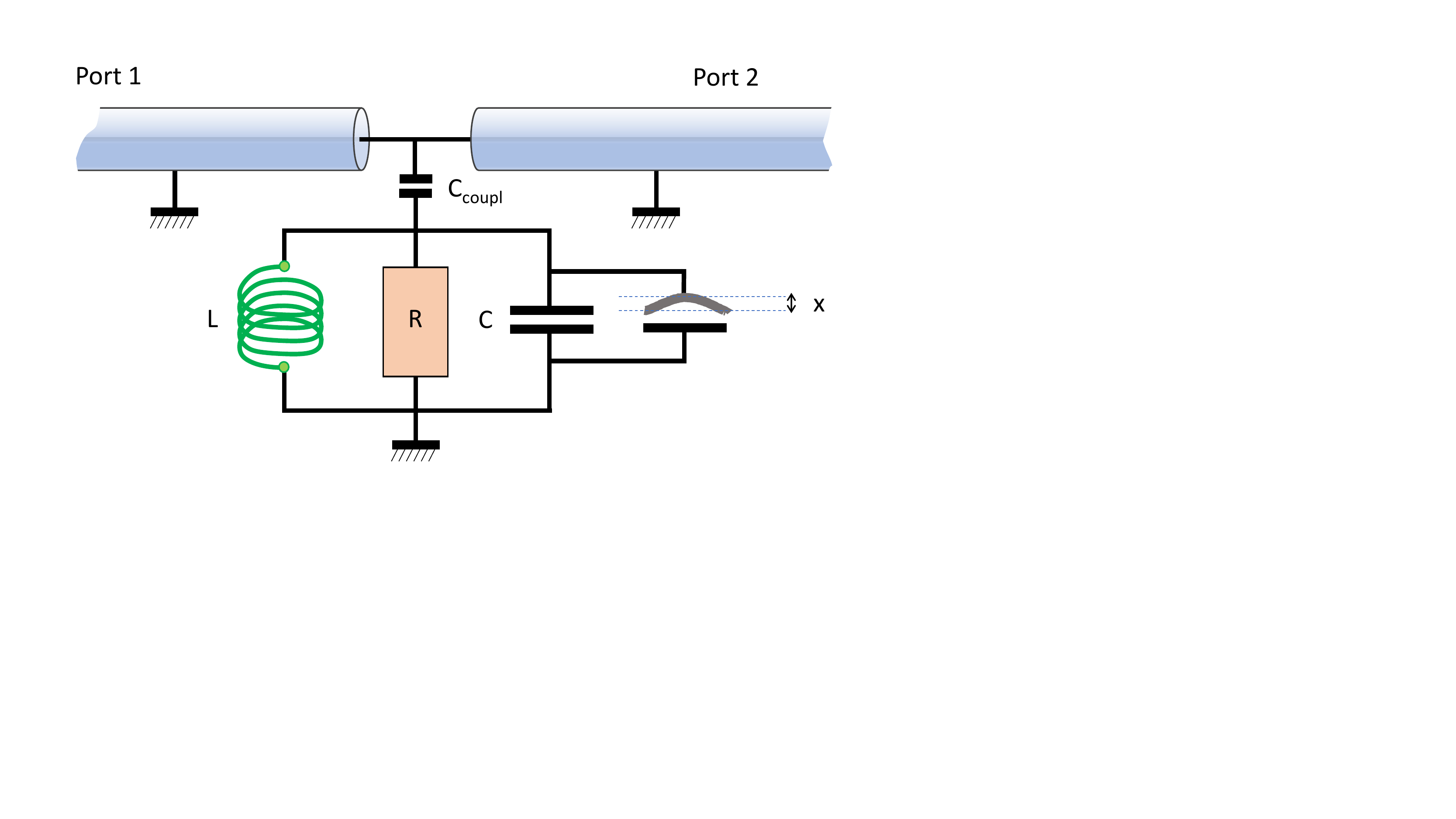}
	\vspace*{-3.5cm}
			\caption{
			(Color online) Schematic of a simple microwave $RLC$ optomechanics circuit. Two-port configuration, similar to Ref. \cite{classic} (see text).}
			\label{fig_0}
		\end{figure}

Beyond the various statistical and spectral properties which have been proposed for it, the stochastic field itself is always treated as a {\it classical} variable (because it is mathematically the simplest), 
while Nature would certainly have it quantum.  
These refinements are outside of the scope of the present manuscript, which is dealing with the possible implementation of experiments looking for {\it quantum-mechanical decoherence and collapse}. 
The point here is that {\it all} Collapse theories propose that the localization lengthscale $\lambda_C$ has to be of the order of $100 \,$nm, right within the {\it mesoscopic range of motion} attainable with centre-of-mass motions of micro-and-nano mechanical systems (MEMS and NEMS) \cite{bassiReview}. 
Centre-of-mass motion on the scale of $\lambda_C$ is key for sensing gravitational fields; but note that the postulated $\lambda_C$ has actually nothing to do with the (impressively small) Planck scale $\sqrt{\hbar G/c^3} \approx 1.6 \times 10^{-35}\,$m \cite{bassiReview}.
Making experiments on some type of MEMS/NEMS superposed quantum state with an amplitude of motion crossing the $\lambda_C$ value, it should be thus possible to directly probe these theoretical proposals. Note that this essentially excludes GHz modes, which display very small motion amplitudes (or even zero centre-of-mass displacements) \cite{cleland2010,clelandSAW,DelsingSAW,entangleSimon,painterGND}. \\

In the following Sections of this Article, we shall first describe the ideal platform (in our view) for research focused on these 3 aspects ({\it  Thermal properties - Material Science - Foundations of QM}), and then give a few hints about experiments that could be conducted there with today's technology. And which obviously {\it could not} be done on any other platform.

\section{Experimental platform: identifying the baths}

Nanomechanics gives us a unique tool to study heat transport, stochastic thermodynamics, and related aspects of material science and QM-fundamentals, down (in principle) to {\it a single phonon mode}. 
We are talking here about rather big objects (with one dimension in the 10 micron range), flexible enough to be able to move substantially (typically with amplitude up to about a micron), in order to be compliant with all aspects discussed in the previous Section.
What is in the focus is suspended top-down fabricated objects with typical thickness {\it large compared to the size of an atom}, typically about 100$\,$nm. We shall not comment on bottom-up structures (nanotubes, graphene, MoS$_2$) which are certainly adapted for some of the topics described in the Introduction, but not for all (especially the centre-of-mass displacement issue of CSL tests, where decoherence directly depends on the transverse dimension to the motion \cite{bassiReview}). 
The actual geometry of the device is not a matter of discussion, it can be a beam or a drum or anything alike; but the constraint 
 on available materials and size shall fix the resonance frequency of the lowest (flexural) mode around $5-50\,$MHz, for typical zero-point-fluctuations $x_{zpf}$ of order $10~$fm. \\

{\it The platform - }
The traveling phonons describing the bulk material are now confined in standing waves within the structure, where a single (added, removed) energy quantum is a {\it collective motion} of all the suspended device.
If the moving object is well-coupled to a (quantum-limited) detector, the sensitivity is strongly enhanced compared to ``catching'' a single quasi-particle in a traveling wave.
We thus consider studying one (or a few) mechanical mode(s) of the structure. Potentially the lowest, for which the thermal population is the largest.
The {\it quantum ground state} will simply be defined by a thermal population $n_{th}$ for the mode lower than 1 on average. Shall the lowest mode verify this inequality, {\it all} the higher modes will be in their quantum ground states.
This is conceptually fundamentally different from experiments studying a very high (GHz) frequency mode
\cite{cleland2010,clelandSAW,DelsingSAW,entangleSimon,painterGND}, or actively cooling only the lowest frequency one \cite{TeufelGrnd,aspelGND}:
In the present case, {\it the whole system} is in its quantum ground state, which leads to the possibilities discussed in the Prospects Section.

What one needs then is a fairly good detector for the motion of a mode (enabling in the first place amplitude-and-phase detection): {\it optomechanics} is perfectly suited for the task \cite{braginsky,caves}. This has been strikingly demonstrated by the detection, as originally suggested, of gravitational waves \cite{ligo}. 
Using microwaves makes optomechanics perfectly compliant with cryogenics \cite{regal2008}: photons are much less energetic, and spurious heatings due to absorption in the materials can be constrained.
The detection of microwave signals is then performed by means of cryogenic ultra-low noise High Electron Mobility Transistors (HEMTs), combined with (almost) quantum-limited amplifiers: Josephson Parametric Amplifiers (JPAs), or nowadays Traveling Wave Parametric Amplifiers (TWPAs). 
Such setups have been demonstrated to be essentially at the Standard Quantum Limit (SQL) for motion detection \cite{LehnertSQL}, which is why we can qualify them as being {\it quantum-limited}.
For a continuous phase-and-amplitude detection, SQL means that the first stage of quantum amplification can reach the minimum back-action allowed by QM: for a measurement performed with photons at frequency $\omega_c$, it requires to feed-back noise to the system with an amplitude equivalent to a half quantum of energy $ \hbar \omega_c$ \cite{clerkPRAmplif,inputoutput}. Practical devices realized today have specifications within 1-3 SQLs, which is why we used the words ``almost'' or ``essentially'' in the above \cite{nico}; we shall not comment on this practical limitation in the following, and will just refer to these devices as being at the quantum limit.

In a microwave optomechanics circuit, the ``optical cavity'' is replaced by an $RLC$ circuit \cite{regal2008,aspelmeyerreview}, with typical resonance frequency around 5$\,$GHz.
The moving mirror is then a deformable capacitor $C$, which we write $C(x)$, see Fig. \ref{fig_0}. 
$x$ is the motion amplitude of a given mode, defined as the amplitude of motion at maximum of the distortion. 
This $RLC$ circuit is coupled to microwave coaxial lines enabling to feed energy in, and recover the signal out. Conceptually, the simplest setup is the 2-port configuration, with evanescent coupling from the microwave cavity to a traveling wave, here through a capacitor $C_{coupl}$; similar setups can of course be built using laser optics \cite{favero}. 
Such a configuration enables a straightforward separation of signals in and out (which is particularly useful for calibration purposes). But it comes at a cost: half of the photons coming out of the device go into the incoming port, rather than towards the output, and are therefore lost.
This can be overcome by using a reflection scheme, where signals in and out are separated via a non-reciprocal element (e.g. a circulator) \cite{aspelmeyerreview}. A purely {\it classical} electric description of such setups can be found in Ref. \cite{classic}.

The $RLC$ resonator essentially acts as a (quantum) {\it transducer}, converting phonons (in the MHz range) into photons (in the GHz range). From the parametric coupling between the $RLC$ and mechanics, the motion imprints sidebands in the microwave signal coming out of the cavity.
Within these setups, various schemes have been developed; in particular, they can be used as {\it non-invasive probes} for motion detection \cite{aspelmeyerreview}.
Technically, this requires to probe the cavity at low enough powers (avoiding heating and other spurious power-dependent effects), for instance by applying a pump tone at $\omega_c$ and measuring the microwave signal spectrum coming out around $\omega_c \pm \omega_m$ (with $\omega_c$ the cavity resonance frequency and $\omega_m$ the mechanical mode frequency). 
Other schemes have been developed, allowing to cool down or amplify a mode's Brownian motion; see e.g. Ref. \cite{aspelmeyerreview} for a review. 
A remarkable class of schemes should be mentioned: Back-Action Evading (BAE) measurements, which allow to {\it ``beat''} the SQL. 
What is actually meant here, is that such measurements allow to measure one quadrature of the motion (say $x$) while feeding {\it all the back-action noise} to the other one (i.e. $p$). As such, $x$ can be measured with an arbitrary precision, at the cost of a complete loss of information on $p$. This is realized in optomechanics by applying two symmetric pump tones, at both $\omega_c \pm \omega_m$, while measuring the spectrum at $\omega_c$ \cite{braginsky,clerkNJP,SchwabBAE}.
Even more complex schemes can be produced, where an {\it effective} mechanical mode (i.e. created from the combination of actual modes) can be essentially free of back-action, see Ref. \cite{laureScience}. This is outside of the scope of our discussion.

		\begin{figure}
		\centering
		\vspace*{-0.5cm}
	\includegraphics[width=15.5cm]{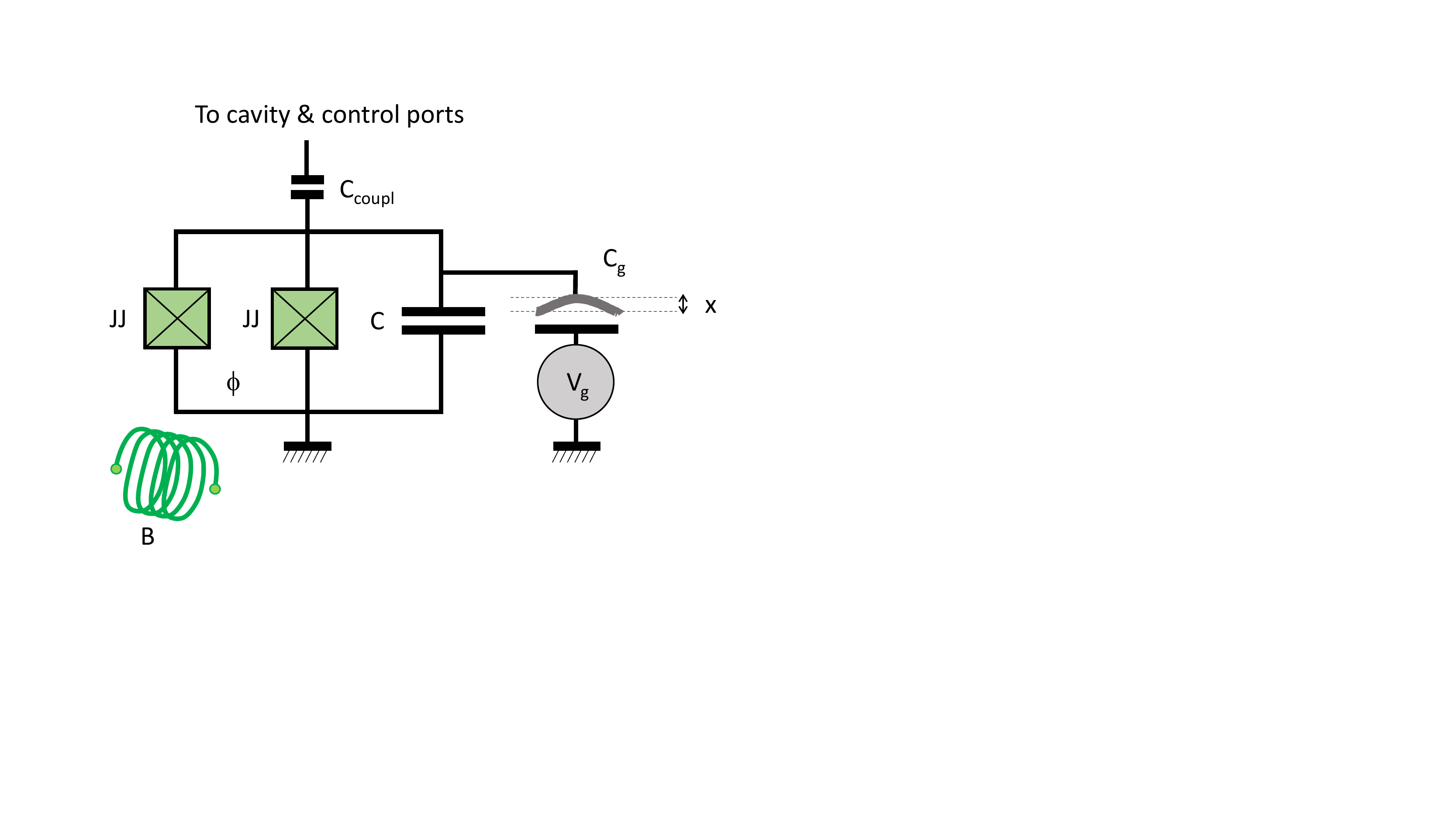}
	\vspace*{-3.5cm}
			\caption{
			(Color online) Schematic of a simple quantum bit-mechanics capacitive coupling circuit, with microwave readout operated through $C_{coupl}$ (not represented). Similar to Ref. \cite{MikaFock} (see text).}
			\label{fig_01}
		\end{figure}

Having the ability to detect the motion, one needs also a way to drive (and tune) the mechanics {\it classically}. This can be achieved (to some extent) by optomechanics alone. One can drive a so-called coherent motion (i.e. a ``classical'' oscillation) by means of a phase-modulated pump \cite{seiscapelle}, or by using the amplifying scheme (a strong microwave pump detuned at $\omega_c+\omega_m$) beyond the parametric instability, creating self-sustained oscillations \cite{selfoscill}.
An even more versatile solution is to add an extra DC+RF port to the mechanical system. This gives the ability not only to drive any specific mode, but also to tune their frequencies \cite{xindrum}: one can adjust them with a DC signal, but also implement (purely mechanical) parametric schemes, which can in turn be used to control the microwave signals \cite{steeleGate}. Such extra gates are not represented in Figs. \ref{fig_0} and \ref{fig_01}, and will not be further described.

Furthermore, in order to study the decoherence of non-classical mechanical states, one needs in the first place to {\it create} such QM states.
To do so, one has to interface the mechanical mode with a quantum object, say a spin-like system; indeed, what is more quantum than that?
Building on the compatibility of microwave optomechanics with quantum electronics, this can be 
 a superconducting qubit, a SQUID-like loop with two Josephson junctions (JJ) shunted by a capacitor, a so-called {\it Transmon} (with particularly long coherence times $t_2$) \cite{qubit,T2transmon}. 
The qubit state is tuned by means of an applied flux $\phi$ through the SQUID loop (via a tiny field $B$), and a gate charge biased through a voltage $V_g$.
It is this gate capacitance $C_g(x)$ which depends on the motion amplitude $x$ of the mechanics \cite{ViennotFockState,MikaFock}; a simple example is shown in Fig. \ref{fig_01}. 
The qubit is also coupled to incoming/outgoing microwave ports in order to prepare and read-out its quantum state. 
As for optomechanics, this is done with a capacitive coupling $C_{coupl}$ to a cavity and coaxial lines (not shown on figure); see e.g. Ref. \cite{qubit}.

A direct capacitive coupling enables to create mechanical {\it  Fock} states (eigenstates of the phonon number) when the qubit is resonant with the mechanics, by direct exchange of quanta through qubit-mechanics {\it Rabi oscillations} \cite{cleland2010}.
This is not adapted for MHz mechanical modes, since Transmon qubits are working at GHz frequencies. 
Conversely, the qubit-mechanics coupling described above is essentially dispersive, meaning that the energy splitting $\hbar \omega_Q$ between the two states $\left|0\right\rangle$ and $\left|1\right\rangle$ of the qubit depends on the motion amplitude $x$; the reverse being also true, i.e. the mechanical frequency $\omega_m$ depends on the qubit state \cite{ViennotFockState,MikaFock}. 
The coupling strength is proportional to $V_g$, and the interaction can thus be switched on/off at will.
It turns out that Fock states can also be created with such non-resonant qubit coupling using a sideband-pumping scheme \cite{MikaFock}, directly adapted from trapped ions physics \cite{ions}. 
It is also technically very similar to optomechanics amplifying and cooling schemes, which share the same origin \cite{aspelmeyerreview}.
One applies a pump tone 
at frequency $\omega_Q \pm \omega_m$ to the combined qubit-mechanics system. 
The interaction is then equivalent to a scattering event in optics:  
the pump photon combines with an excitation from the mechanical mode and excites/de-excites the qubit, while conversely the mechanics is de-exited/excited \cite{MikaFock}.
With these schemes, one can transfer energy by single quanta, which is obviously not possible with a classical drive.

Other types of non-classical states can be achieved with a dispersive qubit coupling \cite{ArmourSchwab,armourNJP}.
Since the mechanical mode oscillation frequency depends on the qubit state, when the latter is put into a superposition state $(\left|0\right\rangle+\left|1\right\rangle)/\sqrt{2}$, it gets entangled with the former in such a way that two ``classical'' motions (i.e. coherent states) with slightly different frequencies are superposed. This means in particular that at specific times within the evolution of the dynamical state, the object will be present at two physically separated positions in space! 
This is particularly neat because one can create such an entangled qubit-mechanics state with arbitrary motion amplitude; one could call it a {\it mechanical Schr\"odinger cat state}, in analogy with what has been already realized with electromagnetic systems \cite{grangier}.
We shall not discuss entanglement of distinct mechanical modes \cite{entangleMika,entangleSimon}; they strikingly tackle these counter-intuitive aspects of quantum mechanics, but their complexity is beyond the scope of this Article.

There is certainly a technical challenge in combining together optomechanics (Fig. \ref{fig_0}), quantum bit (Fig. \ref{fig_01}, which can nonetheless share the same microwave addressing ports), and DC+RF lines within one experiment. Such a task relies on the cleverness of experimentalists, and we do not discuss the issue any further.
But somehow, this arrangement is the ``simplest'' platform that can address {\it all} the points discussed in the Introduction. 
Provided the lowest achievable temperatures for solid state physics come into play.
Indeed, for the microwave cavity and the qubit (with characteristic frequencies around 5$\,$GHz), the required temperatures to guarantee vanishingly small thermal populations (and thus proper quantum-limited manipulation) are below typically 100$\,$mK; which is commercially available using {\it dilution cryostats}.
It should be pointed out at that stage that cooling down electrons in mesoscopic structures to temperatures of the order of 10$~$mK (or even below) is an extremely difficult task \cite{electronT}. This is due primarily to the electron-phonon coupling that falls very rapidly with decreasing temperature, which leads to a {\it thermal decoupling} of the electronic bath for vanishingly small heat inputs. This is obviously not an issue for mechanics.
While it is in principle possible to create superconducting qubits with frequencies resonant with the mechanical modes we consider here (e.g. {\it Fluxonium} qubits, with very good coherence times \cite{fluxo}), these would be extremely difficult to operate because of the required working temperature. This is why we only consider here the dispersive coupling with a GHz Transmon.
Indeed, the mechanical modes under study have resonance frequencies of the order of 10$\,$MHz: passive quantum ground state cooling is thus achieved for sub-milliKelvin temperatures, ensuring one cools the mechanical mode {\it and} the baths that couples to it. This can only be reached using (not commercially available) {\it nuclear demagnetization cryogenics}.
Note however that for kHz devices (like e.g. levitating microspheres \cite{aspelGND} or soft cantilevers \cite{vinante}), the required temperatures are to date still beyond reach of this cooling technique.
The prospects discussed in the following are thus based on the use of this ultimate cooling technology together with microwave quantum techniques; for further discussions on ultra-low temperature topical issues, we refer the reader to Ref. \cite{emp}. \\

\begin{figure}
		\centering
		\vspace*{0.5cm}
		\hspace*{-0.5cm}
	\includegraphics[width=11cm]{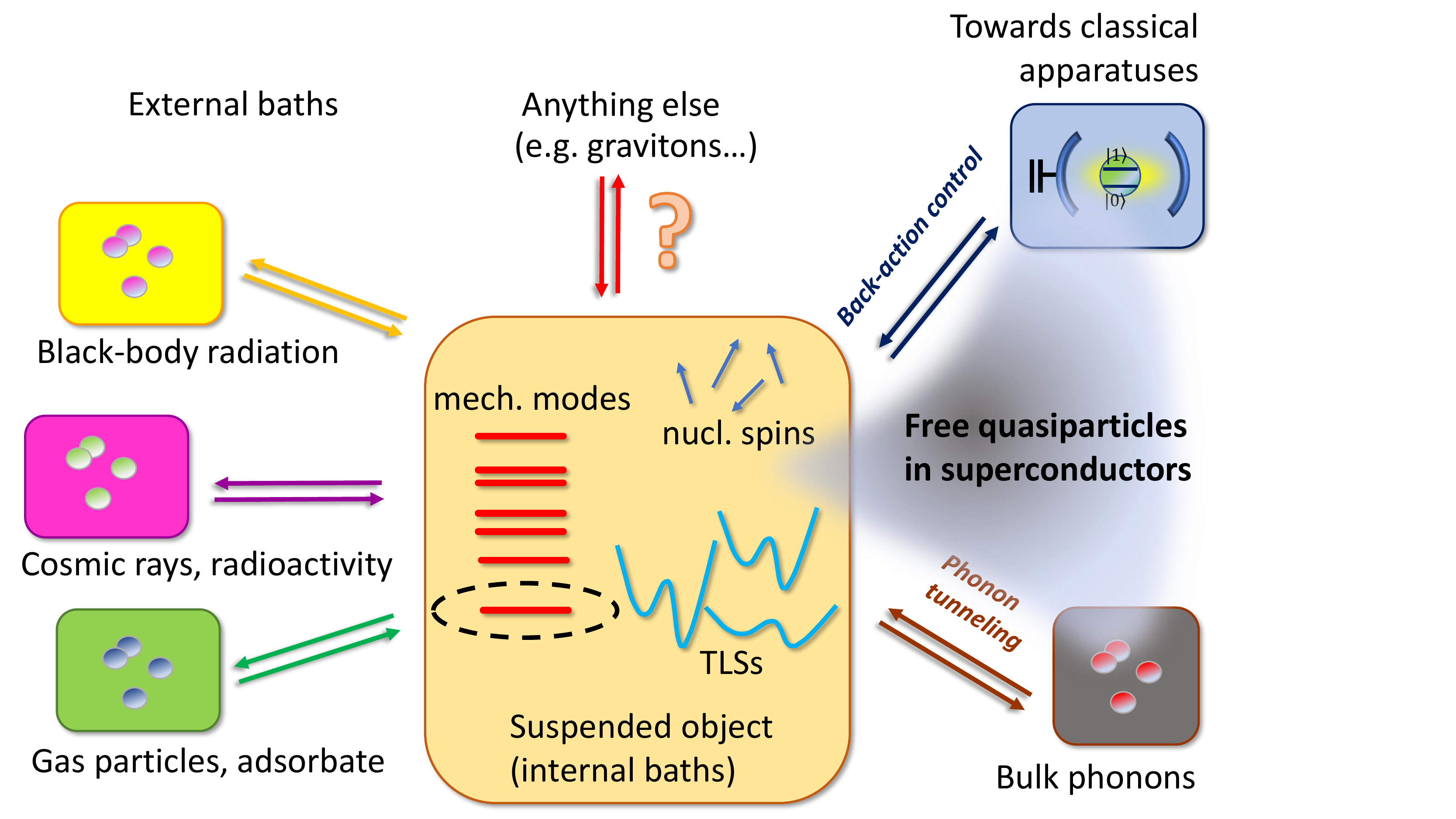}
	\vspace*{0.3cm}
			\caption{
			(Color online) Schematic of the identified baths in contact with the system under study defined as one (potentially the lowest, dashed circle) mechanical mode of a moving mesoscopic structure. See text for details.}
			\label{fig_1}
		\end{figure}

{\it Thermodynamic baths - }
What we want to point out here is that the most interesting in this business is not the coupled (quantum) dynamics of a (few) electromagnetic and mechanical modes, but really the study of the {\it underlying baths}.
And if one wants to detect new contributions (like the postulated noise source of CSL theories), the first step is to clearly identify the known baths coupled to the mechanical mode(s) under study. This is done in Fig. \ref{fig_1}.
Importantly, with a suspended object one can distinguish degrees of freedom {\it inside} the mechanical element from those {\it outside};
this is not possible for e.g. Surface-Acoustic-Wave (SAW) devices (which are not suspended) or levitating microspheres (which have no clamping points). 
This physical separation is a key point if one wants to carefully model phonon tunneling \cite{clamps} (and thus quantum thermal transport) in-and-out one of the mechanical modes. 
Besides, the ensemble of all other mechanical degrees of freedom of the suspended structure constitutes a {\it specific bath}: which is nothing but confined phonons.
This bath interacts (dispersively) with the mode measured through nonlinear coupling \cite{KunalNlinCoupl}. 
If all these other modes are kept hot while only the one measured is actively cooled to its ground state, their Brownian motion shall affect its coherence \cite{nonlinCoupl,DykmanDecohModeNlin}. On the other hand if they are all in their ground states, their only impact should simply be a renormalization of the studied mode's resonance frequency; a mechanical analogue of the {\it Lamb shift} in atomic physics, already discussed in the framework of a qubit-mechanics hybrid system \cite{lamb,lambII}.  

In the centre of Fig. \ref{fig_1} is represented the (suspended) mechanical device, which we assume to be made of ``conventional'' materials: non-magnetic dielectrics (e.g. Silicon Nitride) and/or BCS-type superconductors (e.g. Aluminium) cooled well below $T_C$ (in zero magnetic field, i.e. free of vortices). 
We have there the mechanical mode under study (say, the lowest) and then all the others.
Obviously, one can study more than one mode at a time, provided their respective couplings to the detection scheme is good enough. 
Measuring two modes leads to specific possibilities discussed below in the last Section, but shall not be addressed any more here.
Within the mechanical device, we also have {\it other} internal degrees of freedom: TLSs, and nuclear spins. 
At the top-right is depicted the drive/detection scheme (which is external to the device; and connected to the ``classical world''). It is supposed to be perfectly controlled, in the sense that it reaches the SQL and one can control the amount of (classical and quantum) noise that is fed back into the mechanics by back-action, at the best allowed by the schemes at our disposal.
Note that this noise arises from {\it a single} mode: the one of the measuring apparatus. It is therefore not a true bath in itself. 
However, a given amount of classical noise mimics a bath at temperature $T$; one can use this property to artificially ``heat'' a microwave cavity mode \cite{ilya}. Besides, ``shaping'' the properties of this artificial bath in order to obtain specific properties (down to the quantum limit) is named {\it bath engineering} \cite{zoller,bathengineer}, which is the concept behind many quantum-limited schemes \cite{TeufelGrnd,clerkNJP,kippenberg,fink,clerkbath}.
 
At the bottom-right of Fig. \ref{fig_1}  one finds the thermal bath of (bulk) phonons that ultimately cools the mechanics (and all other internal degrees of freedom), connected through anchoring points. 
This bath is supposed to have a black-body type spectrum, centered on a frequency which depends on temperature. The mean-free-path of the phonons grows as one cools, and can be as large as a centimetre around a Kelvin. Basics of phonon properties in mesoscopic structures can be found e.g. in Refs. \cite{clelandbook,ziman}. 
These are crucial for the understanding of the clamping points of the structure, thermal gradients, and thermalisation of the different degrees of freedom.
In between the suspended object, the detection system and the phonon bath, one might have free Quasi-Particles (QP) within the superconducting parts of the circuit. 
These are thermal QP at a temperature $T_e$ (which might be different from the cryostat temperature $T$), but also strongly {\it out-of-equilibrium} electrons that ``poison'' the almost empty superconductor; a  phenomenon known in quantum electronics and qubits \cite{QPpoison}.
At the bottom-left are represented all other external degrees of freedom: gas particles still present in the very good vacuum of cryogenic chambers (including adsorbates, i.e. particles that stick onto the surfaces of the device), thermal radiation (which we suppose to come from a screen encompassing the whole system, at the temperature of the phonon bath), cosmic rays and natural radioactivity.
The effect of $\gamma$ radiation has been demonstrated on torsional modes of a large glass pendulum using shielding \cite{radiationJeevak}; this has never been performed with MEMS/NEMS so far to our knowledge.

But this might not be the full story; at the top-centre, the last schematized bath holds for {\it any other} (unknown) external source; 
 for instance gravitons, or any other CSL-postulated noise intrinsic to Nature.
Other types of mechanisms might be invoked, like e.g. the propagation of cracks in the materials under stress, even at low temperatures \cite{crackciliberto}.
What is represented in Fig. \ref{fig_1} corresponds in this sense to a minimalist description of what cannot be avoided in a realistic experimental realization.
Each of these baths has in principle a specific statistics and (effective) temperature, and they might not all be in equilibrium (like e.g. the free QPs).
In an experiment, the actual temperature of each bath deserves to be defined, and demonstrating thermal equilibrium is an issue on its own \cite{dylan}.
Besides, the relevance of most of the contributions depicted in  Fig. \ref{fig_1} (excluding bulk phonons and the addressing port) is {\it unknown}. This is only one of the experimental issues that underlie the following discussion.

\section{Prospects} 

In this final Section, we want to discuss explicitly some experiments that can be realized on such a mesoscopic mechanics microwave/microKelvin platform (with no intention to be exhaustive).
Summarizing the above description, thermodynamics is essentially our {\it basic tool} that enables to tackle the questions presented in the Introduction.
The first step is thus certainly to understand it, down to the quantum regime, with all the complexity of its various inter-connected baths. Which is certainly not a simple task. \\

{\it Thermodynamic ensemble average - }
The first aspect to be commented is obviously {\it mean values} of thermodynamic properties.
This includes the demonstration of ground-state cooling of a mesoscopic device, in-equilibrium with its environment. It requires the definition of the most relevant baths thermodynamic temperatures, together with true steady-state quantum signatures measurements.
This has been strikingly realized in a recent publication, opening the experimental field discussed in the present paper \cite{dylan}: the temperature of the phonon bath (i.e. cryostat), of the internal TLSs, and of the lowest flexural mode of an Al-drum have been reported, demonstrating {\it sideband-asymmetry} in the optomechanics spectrum at the lowest temperatures.
When both the cavity mode {\it and} the mechanical mode are almost empty (i.e. in their ground state), sideband-asymmetry is visible as an imbalance of exactly 1 quantum between the measured populations of the two motional sidebands imprinted in the optical spectrum \cite{sidebandassym}. This is a purely quantum signature, which has a classical analogue when classical (microwave) noise is fed back to the mechanics \cite{ilya}.
In Ref. \cite{sidebandassym}, ground state cooling of a MHz mode is obtained by active (sideband) cooling, while in Ref. \cite{dylan} this is performed (for the first time in such systems to our knowledge) by passive cooling.
The key challenge in this experiment is indeed the cooling technology: with a mechanical mode resonating at 15$~$MHz, the lowest temperature achieved to guarantee ground-state cooling was 500$~\mu$K \cite{dylan}.
Ultra-low temperature is one of the frontiers of Science, which is solicited today in particular by modern research in quantum materials and quantum technologies \cite{emp}. The work of Ref. \cite{dylan} is thus part of a European Infrastructure called {\it European Microkelvin Platform} (see acknowledgment).

Beyond pure equilibrium, heat transport can be studied down to the quantum regime. Ideally, under small thermal gradients ensuring that it represents only a (vanishingly) small perturbation of the equilibrium condition \cite{oliveconduct}.
Such experiments can be performed with the monitoring of mechanical modes' effective temperatures while a heat current is imposed, similarly to what is done using cantilevers subject to lasers \cite{ludoBellon}. From the mode temperature's knowledge, one can built a thermal model of the system at hand.
This can also be performed with systems (strongly) out-of-equilibrium, and requires then a completely new conceptual approach (a generalized version of the fluctuation-dissipation theorem) \cite{ludoBellon}; this is not in the scope of our discussion here.
In principle, this could be adapted to the microwave/microKelvin platform we describe here, down to the quantum regime. Either by relying on microwave absorption for the heating, or designing specifically a setup where a heat gradient can be imposed by some other means.

Building on the ability to control equilibrium at temperature $T$, one can study steady-state properties of driven {\it non-classical states}. 
This reduces in the first place to reproducing beautiful experiments already realized under active cooling, but now with an environment that is also cold, thus characterizing {\it the impact} of this environment on the quantum properties. One can immediately think about Fock states creation \cite{MikaFock}, or entanglement \cite{entangleMika} (which can also be obtained with two modes within the same structure, in principle, sharing thus the very same local baths).
Besides, some exciting proposals which require such ground-state conditions have not yet been realized, 
precisely because the experimental conditions were not available yet. For instance, with a very cold environment, the self-oscillating state (triggered with the optomechanical blue-detuned scheme beyond threshold) can display extremely low amplitude (and phase) noise, with statistics that can become {\it non-Gaussian} \cite{ArmourPRL}.
Also, some proposed optomechanical tests of quantum gravity (that aim at probing Planck scale corrections, which we shall not discuss here) require the knowledge of the Hamiltonian's nonlinearities \cite{nonlinhamilton}. 
Such nonlinear terms can be obtained experimentally by a careful fitting of the response obtained in the self-oscillating regime \cite{selfoscill}. With microwave setups, it concerns the higher-order coefficients of the Taylor expansion of $C(x)$ [``conventional otpomechanics'' being built around the first linear coupling term], plus the intrinsic Duffing nonlinearity of the mechanical mode (frequency-pulling arising from stretching).
These nonlinearities are mandatory as well for the proper modeling of very large amplitude of motion response. 
The most speculative experiments addressing directly macroscopic-states collapse, presented in the very last part of this Section, would thus require this knowledge in the first place. 

CSL and gravity-related Collapse models predict an ``effective'' thermal decoupling of mechanical modes at $T \rightarrow 0~$K due to the random field introduced. 
This is a purely {\it classical} effect, which depending on the actual values of $\lambda_C$ and $\Gamma_C$ could be measurable \cite{diosi,vinante2}.
It obviously requires to master the measurement of $T$, and the demonstration that there is no instrumental thermal saturation.
Experiments are conducted (at higher temperatures) trying to set bounds on this effect, using soft cantilevers \cite{vinante} and levitated microspheres \cite{barker}. MHz mechanical modes are not really suited for these studies \cite{diosi,vinante}, but in principle a microKelvin platform can host other types of devices. \\

{\it Thermodynamic fluctuations - }
Quoting R. Landauer: ``the noise is the signal'' \cite{landauernoise}.
And indeed there is a lot of information in the {\it fluctuations} around the  mean of thermodynamic quantities that deserve to be carefully analyzed.
Mesoscopic mechanics is also particularly well adapted to such measurements because mechanical relaxation times $t_m$ can easily be in the range 1 - 100$~$ms for MHz modes \cite{NEMSdamping}. This is well within the bandwidth of the GHz detection chain; and this time-resolution can be met provided the first amplification stage meets optimal noise specifications.
In principle, it should be possible to measure the imprint of {\it single-phonon tunneling} in-and-out a mechanical mode, 
 and link it to generic thermal transport properties. 
One would require to measure the mechanical resonance peak in the motion spectrum at a fast enough speed, constrained by the limit of the mechanical decay rate itself.
Such motivations were discussed already in the literature from both an experimentalist's \cite{thermalROUKES} and a theoretician's \cite{thermalMILES} perspective.
Within an idealized system, the mechanical decay time would indeed be dominated by clamping losses at the lowest temperatures, and the measured population spectrum should reproduce a simple {\it Ornstein-Uhlenbeck}  process (a flat noise up to a frequency $\sim 1/t_m$) \cite{OUprocess}.
Realizing the measurement down to the lowest temperatures might clarify if indeed a quantum correction appears in the quantum version of the fluctuation-dissipation theorem applied to heat \cite{averinjukka}.

Mechanical properties themselves are fluctuating, namely the mode resonance frequency and damping \cite{natnanotechsansa,oliveNoise,dylan}.
Their temperature dependencies measured on an Al-drum below 100$~$mK down to 500$~\mu$K are reminiscent of what has been reported for superconducting cavities \cite{natcommTLS}. In the latter case, these properties have been attributed to TLSs present within the superconductor, and a model of interacting TLSs was fit to the data. For mechanics, we can only speculate that these very-low $T$ mechanical fluctuations have the same intrinsic origin; no model has been fit yet on data.
Besides, an increase in mechanical damping is expected at the lowest temperatures, due to the so-called resonant-TLS contribution \cite{TLS}.
This is not seen in the measurements, and one can argue that the reason is the {\it saturation} of TLSs with extremely low levels of strain \cite{remus}. 
Further experiments are clearly required to clarify the issue. 
In this respect, it has been proposed to directly probe individual TLSs by coupling them with a ground-sate cooled mechanical mode \cite{AFTLS}.
In a similar fashion to what has been achieved with qubits and an electromagnetic addressing \cite{ustinovTLS}, it could thus be possible to measure the density of TLSs, their relaxation and coherence rates, relying {\it only} on the strain coupling. 
More complex schemes could even be envisaged, like phonon echoes \cite{phononecho}, implying a {\it single} TLS entity.
The ultimate goal being to identify these microscopic entities and control them at will.

Finally, the reported measurements on low temperature microwave optomechancial setups already present {\it puzzling stochastic features} that are not understood.
Large amplitude ``spikes'' are seen in the motion spectrum below about 100$~$mK for beam devices \cite{zhouPRAppl}.
It seems that a similar instability can be found in drum devices below 20$~$mK, see Suppl. Material of Ref. \cite{dylan}.
The phenomenon seems to be linked to the presence of the microwave drive, but no clear dependence could be defined. 
The mechanism behind it remains a mystery, and more work (both theoretical and experimental) is needed.
Besides, very slow fluctuations of the thermal population have been reported in Ref. \cite{dylan}.
These have a striking temperature-dependence that goes as the square-root of the thermal population. 
This is certainly {\it not} what one would expect from a conventional thermal bath. Again, the origin of this effect remains mysterious.
Beyond entanglement and QM-related experiments, one type of measurements that can be suggested 
is to track {\it correlations} between mechanical properties of different modes of the same structure. 
This could help identifying the origin of these out-of-equilibrium events and very slow signatures. \\

{\it Real-time manipulations - }
Certainly the most ambitious, but also the most promising, is to design {\it time-resolved measurements of the dynamics} of non-classical states.
In other words, the point is to probe the QM-classical cross-over {\it in space and time}, being able to measure the dynamics of arbitrary amplitude superposed states (even larger than $\lambda_C$), at a rather fast speed (faster than the collapse time $t_C$ of the state).
For instance, one could implement an ``echo scheme'' as suggested by the Authors of Ref. \cite{armourNJP}, who analyze a setup compatible with the platform presented here. The idea is to create a quantum-coherent superposition of two ``classical'' motion states of arbitrary amplitude $\eta \, x_{zpf}$  (and $\eta \geq 1$) with two slightly different frequencies, thanks to the entanglement with a $(\left|0\right\rangle+\left|1\right\rangle)/\sqrt{2}$ qubit state (see setup description).
The system is then left to evolve freely over a period of time $\tau$, after which the qubit state is ``inverted'': the quantum-coherent superposition then evolves ``backwards'', in such a way that after a time $\tau$ the two ``classical'' states re-focus. One then measures the qubit population in state $\left|1\right\rangle$.
This protocol has already been implemented in a qubit alone (with no coupling to a mechanical mode) \cite{ithier}, and has the specific feature of canceling decoherence from slow fluctuations of the qubit energy levels (a phenomenon also named re-coherence in this framework), leaving only the damping contributions. 
This also enables us here to work with the longest time-frame $t_{echo}> t_2$ for mechanically-induced decoherence measurements.
With nonzero coupling, the mechanics then imprints characteristic {\it oscillations} in the measured re-coherence of the qubit \cite{armourNJP}.
Then the point would be to study how much this pattern depends on  the temperature of the baths, and maybe more importantly the {\it motion amplitude}. The impact of the baths (TLS, phonons, microwave readout) can be calculated; one calls it {\it environmental decoherence} \cite{armourNJP,bouwmeesterNJP,Steeledecoh}. 
With the simple assumption of thermal equilibrium at temperature $T$, a standard 10$~$MHz device with a global quality factor of $Q \sim 5 \times 10^4$, one calculates a typical environmental decoherence time $t_{env} \sim 0.4/\eta^2~$ms around $T\sim 1 ~$mK \cite{bouwmeesterNJP,Steeledecoh}.
This is of the order of the best Transmon $t_{2}$ values for $\eta=1$ (and otherwise shorter) \cite{T2transmon,T2transmonBEST}; 
and definitely shorter than the mechanical decay times $t_m$.
Since this time is shorter, it should be measurable. Will experiments match theory? Can one disentangle the contributions of each baths, even the ones that might stay out-of-equilibrium (as free QPs, for instance)?
On the other hand, the collapse time estimates $t_C$ are extremely speculative: the rates $\Gamma_C$ (for a single nucleon) introduced in Collapse theories vary from typically $10^{-17}~$s$^{-1}$ to $10^{-8}~$s$^{-1}$ \cite{bassiReview}, with scalings for $t_C$ as $n$ or $n^2$  with the number of particles depending on the models (distinguishable or non-distinguishable particles). 
Specific gravitational Collapse models vary also substantially, depending on the typical scales taken to define {\it what mass} shall be actually considered in the decoherence process (the mass of a proton, or of the whole object?) \cite{bassiReview,bouwmeesterNJP,Steeledecoh}. This leads to estimates from well-below a ms up to more than a second.
Besides, does something {\it special}   happen for motion amplitudes reaching about $100~$nm \cite{bassiReview}?
These are exciting questions which will certainly remain unsolved for a few more years.

\section{Conclusion}
In this prospect Article, we introduced the concepts and open questions linked specifically to the mechanical behavior of small systems, down to the quantum regime. The focus is on thermal transport and thermal equilibrium, fluctuations and thermodynamic properties, material science and quantum mechanics foundations.
We discuss a specific class of experimental arrangements, built around three main components: mesoscopic devices that can {\it reasonably move}, quantum-limited 
{\it microwave circuits} with optomechanics and qubits, and {\it ultra-low temperature} cryogenics. 
Some experiments at hand with today's technology are presented.
The point of view that is defended here is that these specific mechanical objects are far more than model systems reproducing expectations from (classical and) quantum mechanics: these are {\it quantum sensors}, and the subject of the research is actually not the mechanical modes themselves, but the {\it thermodynamic baths} to which they couple. 

\section{Data availability}
Data sharing is not applicable to this article as no new data were created or analyzed in this study.

\section{Conflict of interest}
The author has no conflicts to disclose.

\begin{acknowledgements}

The Author wishes to thank extremely valuable discussions with Andrew Armour, Olivier Maillet, Mika Sillanp\"a\"a, Andrew Fefferman, Olivier Bourgeois, Xin Zhou, Fabio Pistolesi  and Henri Godfrin.
He acknowledges support from the ERC CoG grant ULT-NEMS No. 647917. 
This research has received funding from the European Union's Horizon 2020 Research and Innovation Programme, under grant agreement No. 824109, the European Microkelvin Platform (EMP, \underline{https://emplatform.eu/}). This article is dedicated to the memory of Dylan Cattiaux \cite{dylan}, who opened the experimental filed of macroscopic quantum mechanical motion at ultra-low temperatures.

\vspace*{0.1cm}
(\dag) Corresponding Author: eddy.collin$@$neel.cnrs.fr

\end{acknowledgements}


\end{document}